\documentclass[onecolumn,noshowpacs,superscriptaddress,nobibnotes,nofootinbib,12pt]{revtex4}
\usepackage{amsmath,amssymb}
\usepackage{epsfig}
\usepackage{graphicx}
\newcommand{\be}{\begin{equation}}
\newcommand{\ee}{\end{equation}}
\newcommand{\bea}{\begin{eqnarray}}
\newcommand{\eea}{\end{eqnarray}}
\newcommand{\nn}{\nonumber}

\newcommand{\f}{\frac}

\newcommand{\bra}{\langle}
\newcommand{\ket}{\rangle}

\newcommand{\bd}{\mathbf}
\newcommand{\p}{\partial}

\begin{document}
\title{On heavy-quarkonia suppression by final-state multiple scatterings in most central Au~+~Au collisions at RHIC}

\author{Bin Wu}
\email{binwu@physik.uni-bielefeld.de} \affiliation{School of Physics and State Key Laboratory of Nuclear Physics and Technology,
Peking University, Beijing 100871, China}
\affiliation{Faculty of Physics, University of Bielefeld, D-33501 Bielefeld, Germany}

\author{Bo-Qiang Ma}
\affiliation{School of Physics and State Key Laboratory of Nuclear Physics and Technology,
Peking University, Beijing 100871, China}

\begin{abstract}
We study heavy-quarkonia suppression under final-state multiple scatterings in most central Au~+~Au collisions at RHIC energy. We first calculate the survival probability of a heavy quarkonium under multiple scattering in Bjorken's expanding QGP at large $N_c$. Then, we calculate the rapidity dependence of the nuclear modification factor $R_{AA}$ for heavy-quarkonia production by considering final-state multiple scatterings in most central Au~+~Au collisions in a simplified model. In our formula a constant $P_0$ is also introduced to estimate the possible cold nuclear effects. By fitting the data for $J/\Psi$ production in most central Au + Au collisions at $\sqrt{s_{NN}}=200$~GeV at RHIC, we find that the transportation coefficient  $\hat{q}_0\simeq(0.33-0.95)~\mbox{GeV}^2$/fm, and, accordingly, the energy density at $\tau_0$ is $\epsilon_0\simeq(1.39-5.62)~\mbox{GeV}/\mbox{fm}^3$ in perturbative thermal QCD. A better understanding of cold nuclear effects is essential for us to get a more accurate analysis. The small values of the transportation coefficient $\hat{q}_0$ in our estimate are in sharp contrast with those obtained by the analysis of high-$p_T$ hadron spectra in Ref. \cite{Eskola:2004cr}.
\end{abstract}


\maketitle
\newpage

\section{Introduction}
Relativistic heavy-ion collisions at RHIC and the forthcoming LHC  provide the opportunity  to study QCD matter of extreme temperature and density, the quark-gluon plasma (QGP)\cite{Adcox:2004mh}. Light quarks and gluons are expected to be thermalized to form QGP in a short time $ \tau_0\sim 1~\mbox{fm}$. Heavy quarks are too heavy to be thermalized in such a short time. With a formation time much shorter than $ \tau_0$ they nearly live through all the stages of heavy ion collisions and, in this way, act as an excellent probe to study the properties of the background medium. Because of their small size, heavy quarkonia,
the bound states of heavy quarks, play an important role in understanding the formation and properties of the QCD matter created in heavy ion collisions\cite{Rapp:2009my}.

$J/\Psi$ suppression was originally proposed by Matsui and Satz as a signature for the QGP formation in heavy ion collisions\cite{Matsui:1986dk}: in QGP color screening dissolves $J/\Psi$ into open charmed mesons $D$ and $\bar{D}$. The NA38 collaboration at the CERN SPS observed for the first time the suppression of $J/\Psi$ production in
oxygen-uranium reactions at 200 GeV/nucleon\cite{Baglin:1990iv}. However, the answer to whether $J/\Psi$ suppression can be taken as an unambiguous signature for the QGP formation or not becomes complicated since one can not rule out alternative explanations by cold nuclear effects (see \cite{Satz:1998bd} and references therein).

$J/\Psi$ suppression has been observed with puzzles in Au~+~Au collisions at
$\sqrt{s_{NN}} = 200~\mbox{GeV}$ by the PHENIX experiment at
RHIC\cite{Adare:2006ns}. One puzzle is that the suppression at midrapidty is similar to that observed at the SPS
and smaller than expectations based on the analysis of the local medium density. This possibly implies that the enhancement of $J/\Psi$ production by recombination of
initially uncorrelated $c\bar{c}$ paires are important at RHIC energy\cite{Thews:2005vj,Grandchamp:2005yw}.
The other puzzle lies in the dependence of the $p_T$ integrated
nuclear modification factor $R_{AA}$ on $N_{\mbox{part}}$, the number of participant. $R_{AA}$ is more
suppressed at forward rapidity than at midrapidity. Possible
solutions for such a puzzle have been proposed in
\cite{Kharzeev:2008cv,Zhao:2008pp,Liu:2009wza}.

In this paper, we study the dissociation of heavy quarkonia by final-state multiple scatterings in Bjorken's expanding QGP. At RHIC energy, the heavy quarkonia produced in heavy ion collisions are typically non-relativisitc. Therefore, one can estimate the creation time $\tau_B$ by the inverse of the binding energy $E_B$ in the quarkonium rest frame. If the background medium is thermalized within the thermalization time $\tau_0\sim 1$~fm,  heavy quarkonia will be created before thermalization. Therefore, one can use heavy quarkonia to probe the thermodynamic properties of the system at $\tau_0$ since the information about thermalization can survive final-state multiple scatterings.

$\Upsilon$, the lowest $b\bar{b}$ bound state, is a more ideal probe to decipher information about thermalization of the background medium than $J/\Psi$. $\Upsilon$, with a size $\bra a_B\ket\simeq 0.28~\mbox{fm}$, is the most tightly bound state and the background QCD matter, thermalized or not, is more partonic to such a hard probe. Due to their short formation time $\tau_B\simeq0.18~\mbox{fm}$, $\Upsilon$ can actually witness and record the thermalization of the partonic background medium. Moreover, at RHIC energy the recombination effect is small in $\Upsilon$ production\cite{Grandchamp:2005yw} such that  the information of thermalization carried by $\Upsilon$ can survive final-state interactions. On the experimental side, the study of $\Upsilon$ production in relativistic heavy ion collision for the first time becomes possible\cite{Das:2008zz}. Because of the increased cross-seciont of $\Upsilon$ production at LHC energy, one can expect extensive study of $\Upsilon$ suppression in heavy ion collisions at LHC.

In this paper we focus on the significance of  final-state multiple scattering to heavy-quarkonia suppression and neglect other possibly important effects, such as partonic energy loss\cite{Wiedemann:2009sh}, the gluon saturation effects\cite{Kharzeev:2008cv}, the nuclear absorption and recombination effect\cite{Thews:2005vj,Capella:2007jv}. This paper is organized as follows. In Sec. \ref{sec:ms}, we first give a brief review about Bjorken's hydrodynamic model and a new derivation of the multiple scattering of a quark in it. Then, we calculate the survival probability of a heavy quarkonium in such an expanding QGP at large $N_c$. In Sec. \ref{sec:RHIC}, we investigate the implications of our formula of the survival probability to heavy-quarkonia suppression at RHIC in a simplified model. First, we give an intuitive explanation to the experimental results about $J/\Psi~R_{AA}$ versus $y$ in most central Au~+~Au collisions at $\sqrt{s_{NN}} =200~\mbox{GeV}$. Then, we illustrate in the meanwhile how one can possibly abstract $\hat{q}_0$, the transportation coefficient at $\tau_0$  by studying $J/\Psi$ suppression in Bjorken's
expanding QGP. This quantity is of great significance in our understanding of jet quenching in heavy-ion collisions\cite{Wiedemann:2009sh}. Conclusions are presented in Sec. \ref{sec:con}.

\section{Dissociation of heavy quarkonia by multiple scatterings in an expanding QGP}\label{sec:ms}

\begin{figure}
\begin{center}
\includegraphics[width=10cm]{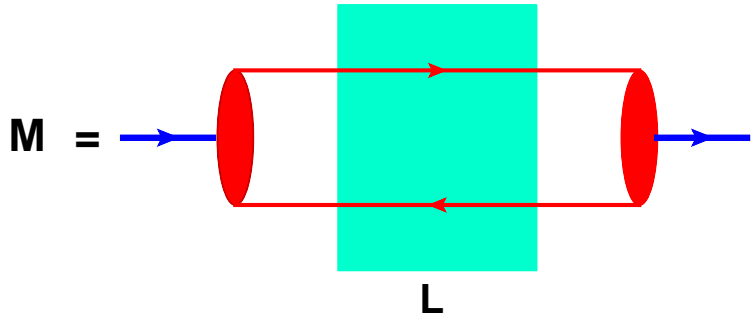}
\end{center}
\caption{Multiple scattering of a heavy quarkonium.}\label{Fig:ms}
\end{figure}

In this section, we give a brief discussion about heavy-quarkonia
dissociation due to final-state multiple scatterings in an expanding
QGP. As illustrated in Fig. \ref{Fig:ms}, a heavy quarkonium with a
size $\bra a_B\ket$, after creation, travels through a slab of expanding QGP
and we will calculate the survival probability for such a process. We use Bjorken's hydrodynamic model\cite{Bjorken:1982qr} to
describe the evolution of the background medium after the thermalization time $\tau_0$.

\subsection{Bjorken's hydrodynamic model\cite{Bjorken:1982qr}}

Imagine that after two gold nuclei collide head-on, the system reaches local thermodynamic
equilibrium at time $\tau_0$. Following Bjorken, we assume cylindrical symmetry in the transverse direction to the collision axis.  The energy density $\epsilon$ and the entropy
density $s$ only depend on the invariant time $\tau \equiv \sqrt{t^2
- z^2}$ and respectively satisfy the following equations

\be \f{d\epsilon}{d\tau} = - \f{\epsilon +
P}{\tau},\label{equ01_Bj}\ee and \be\f{ds}{d\tau} =
-\f{s}{\tau},\label{equ02_BJ}\ee where $P$ is the pressure. In the following, we assume that the pressure and the energy density are related by the following equation
 \be P = c_s^2\epsilon,\ee with the sound velocity $c_s$ a constant.  Inserting into Equ.
(\ref{equ01_Bj}) the following thermodynamic relations

\be \f{dP}{dT} = s~\mbox{and}~\epsilon=Ts -P, \ee

one obtains
\be \f{dT}{d\tau} = -c^2_s\f{T}{\tau}.\ee Accordingly,
one can easily get the so-called Bjorken's scaling solution

\be T = T_0\left(\f{\tau_0}{\tau}\right)^{c_s^2},\label{equ:Ttau}\ee
\be s =s_0\left(\f{\tau_0}{\tau}\right),\ee and \be\epsilon
=\epsilon_0\left(\f{\tau_0}{\tau}\right)^{c_s^2+1},\label{equ:stau}\ee
where $s_0$, $\epsilon_0$ and $ \tau_0$ are respectively the entropy
density, the energy density and the temperature at the thermalization time
$\tau_0$.

The conservation of entropy (see Equ. (\ref{equ02_BJ})) predicts the
central-plateau structure of particle multiplicity in the rapidity
spectra, which is not observed at RHIC\cite{Back:2002wb}. However,
the deviations from boost invariance are not very large for
$|y|\lesssim2.5$. Moreover, our picture about heavy-quarkonia
suppression in the following sections can be easily generalized to
other more complicated hydrodynamic
models.

\subsection{The multiple scattering of a quark in an expanding QGP\cite{Baier:1998yf}}

\begin{figure}
\begin{center}
\includegraphics[width=14cm]{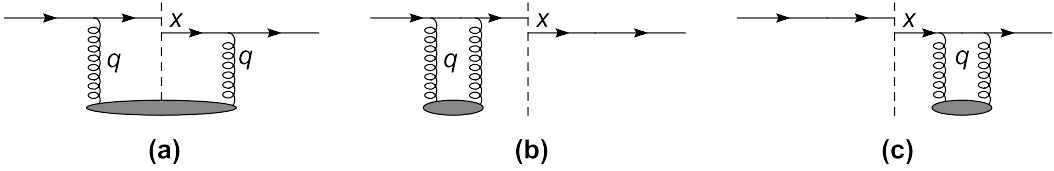}
\end{center}
\caption{One scattering of a quark off the QGP constituents:
$\bd{X}$ is the relative transverse coordinates of the two quark
lines.}\label{Fig_one_scattering}
\end{figure}

When a quark travels in the
 expanding QGP cylinder, it picks up a transverse
momentum from random kicks by the QGP constituents. In the
following, we will give a new derivation of the evolution of
$\f{dN}{d^2p_\perp}(p_\perp,\tau,\tau_0)$, the transverse momentum
distribution of the quark,  from initial invariant time $\tau_0$ up to time $\tau$. The fundamental
assumption here is that the quark undergoes uncorrelated multiple
scatterings with the QGP constituents. Let us define $f(\bd{X},\tau,\tau_0)$
such that \be\f{dN}{d^2p_\perp}(p_\perp,\tau,\tau_0)=\int
d^2xe^{-i\mathbf{p}\cdot\bd{X}}f(\bd{X},\tau,\tau_0).\ee  Given $f(\bd{X},\tau,\tau_0)$,
one can choose an infinitesimal time interval $d\tau$ such that the
difference between $f(\bd{X},\tau+d\tau,\tau_0)$ and $f(\bd{X},\tau,\tau_0)$ can be
calculated by considering only one extra scattering. Including all
the three Feynman diagrams in Fig. \ref{Fig_one_scattering}, in eikonal approximation we have

 \be f(\bd{X},\tau+d\tau,\tau_0)-f(\bd{X},\tau,\tau_0)=f(\bd{X},\tau,\tau_0)d\tau\rho(\tau)
\int\f{d^2q}{(2\pi)^2}\f{d\sigma}{d^2q}(q^2,\tau)\left(e^{i\bd{q}\cdot\bd{X}}-1\right),\ee
or, equivalently,

\be \f{\p}{\p \tau}f(\bd{X},\tau,\tau_0)=\rho(\tau)
f(\bd{X},\tau,\tau_0)\int{d^2q}\f{d\sigma}{d^2q}(q^2,\tau)\left(e^{i\bd{q}\cdot\bd{X}}-1\right),\label{Eq_dfdt}\ee
where $\f{d\sigma}{d^2q}(q^2,\tau)=\alpha_s
C_F\int\f{d^2q}{(2\pi)^2}|V(q^2,\tau)|^2$.

Solving (\ref{Eq_dfdt}) with the initial condition
$f(\bd{X},\tau_0)$, we obtain

\be
f(\bd{X},\tau,\tau_0)=f(\bd{X},\tau_0)\exp\left\{\int_{\tau_0}^\tau d\tau^\prime\rho(\tau^\prime)\int{d^2q}
\f{d\sigma}{d^2q}(q^2,\tau^\prime)\left(e^{i\bd{q}\cdot\bd{X}}-1\right)\right\}.\ee

The transport coefficient $\hat{q}(\tau)$ is defined
as\cite{Baier:1996sk}\be\hat{q}(\tau)=\rho(\tau)\int d^2q
q^2\f{d\sigma}{d^2q}(q^2,\tau).\ee For $X^2\simeq 0$, we
have\cite{Baier:1998yf}
\be
f(\bd{X},\tau,\tau_0)=f(\bd{X},\tau_0)\exp\left\{-\f{1}{4}\bra p^2_{\perp}(\tau,\tau_0)\ket
X^2\right\},
\ee
 where the $p_T$-broadening of the quark is given
by
\be
\bra p^2_{\perp}(\tau,\tau_0)\ket\equiv \int_{\tau_0}^\tau d\tau^\prime
\hat{q}(\tau^\prime).
\ee 
The leading term of the high-temperature
expansion for $\hat{q}(\tau)$ takes the following form
\cite{Baier:1998yf}
\be 
\hat{q}(\tau) = \hat{q}_0\left(\f{\tau_0}{\tau}\right)^{3c^2_s},
\ee
and, accordingly, 
\be
\bra p^2_{\perp}(\tau,\tau_0)\ket = \hat{q}(\tau)\tau\f{1 -
\left(\f{\tau_0}{\tau}\right) ^{1-3c^2_s}}{1 - 3c^2_s}.\label{equ:p2t}
\ee

\subsection{Survival probability of a heavy quarkonium in an expanding QGP}\label{Sur}

\begin{figure}
\begin{center}
\includegraphics[width=14cm]{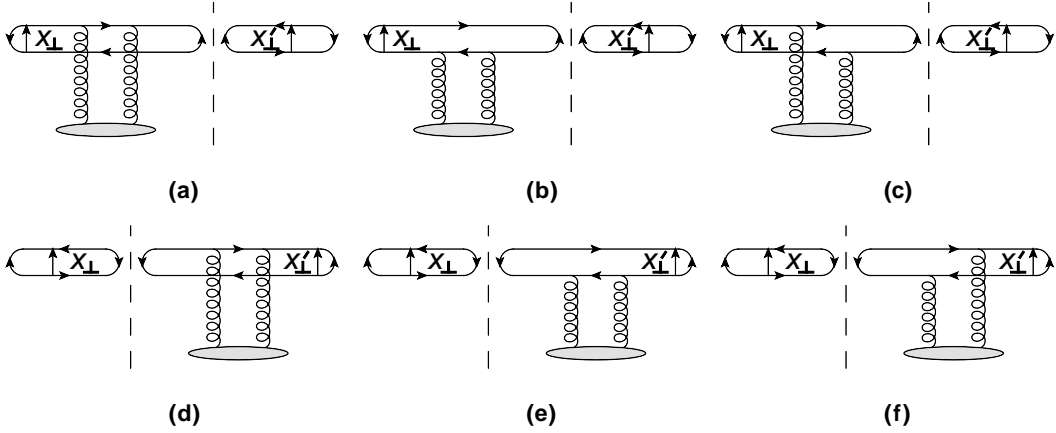}
\end{center}
\caption{One scattering of a color singlet dipole off the QGP constituents: $\bd{x}_\perp$ and
$\bd{x}_\perp^\prime$ are respectively the relative transverse coordinates of the quark and antiquark in the amplitude and in the conjugate amplitude.} \label{Fig:one_scattering_hq}
\end{figure}
After creation, a heavy quarkonium travels through a slab of QGP.
If neglecting higher Fock components, one can take the heavy quarkonium as a $Q\bar{Q}$ color singlet dipole. A detailed discussion of the initial-state nuclear effect and the validity of the dipole model in describing heavy-quarkonia production in heavy-ion collisions is presented in Ref. \cite{Kharzeev:2008cv}. In this paper, we only deal with the dissociation of a color-singlet dipole(quarkonium) after creation, which differs with Ref. \cite{Kharzeev:2008cv}. As illustrated in Fig. \ref{Fig:ms}, the color singlet dipole may break up due to final-state interactions in the QGP. If $\tau$, the time for the quarkonium traveling in the QGP, is smaller than the formation time $\tau_{form}\equiv \gamma \tau_B$, one can neglect the interaction between the dipole and only consider the multiple scattering. However, if $\tau >t_{form}$, one has to consider the mutual interaction between the quark and anti-quark in the quarkonium besides the $p_T$-broadening of them under multiple scattering.  In this subsection, we will calculate the survival probability in these two cases.

\subsubsection{The multiple scattering of a heavy-quarkonium with $t \lesssim t_{form}$}

Within the formation time $\gamma\tau_B\equiv \gamma/E_B$ with $E_B$ the binding energy, one can neglect the interaction between the quark and anti-qaurk in a bound state\cite{Dominguez:2008be}. First, let us calculate the $\tau-$evolution of $f(x_\perp,x_\perp^\prime,\tau,\tau_0)$, the product of the quarkonium state in the amplitude and that in the conjugate amplitude. As shown in Fig. \ref{Fig:one_scattering_hq}, $\bd{x}_\perp$ and $\bd{x}_\perp^\prime$ are respectively the relative transverse coordinates of the quark and antiquark in the amplitude and in the conjugate amplitude. In eikonal approximation, only the relative transverse coordinates of the quark and antiquark are relevant\cite{Dominguez:2008aa}. In the following we will take the light-cone wave function of the quarkonium as a function only of relative transverse coordinates.
Before the initial time $\tau_0$, when there is no QGP present, we have
\be f(x_\perp,x_\perp^\prime,\tau_0)=\varphi(x_\perp)\varphi^*(x_\perp^\prime),\label{Equ:fxx0}\ee with $\varphi(x_\perp)$ the light-cone wave function of the quarkonium.
Neglecting terms proportional to $\mathcal{O}(1/N^2_c)$, for one scattering of the heavy quarkonium off the QGP constituents we need only sum over the six diagrams as shown in Fig. \ref{Fig:one_scattering_hq}\cite{Dominguez:2008aa}.  Following the derivation of $f(\bd{X},\tau,\tau_0)$ for the multiple scattering of a quark in the previous subsection, it is easy to show that in eikonal approximation $ f(x_\perp,x_\perp^\prime,\tau,\tau_0)$ satisfies the following partial differential equation
\be
 \f{\p}{\p \tau} f(x_\perp,x_\perp^\prime,\tau,\tau_0)=\rho(\tau)
 f(x_\perp,x_\perp^\prime,\tau,\tau_0)\int{d^2q}\f{d\sigma}{d^2q}(q^2,\tau)\left(e^{i\bd{q}\cdot\bd{x}_\perp} + e^{-i\bd{q}\cdot\bd{x}_\perp^\prime}-2\right),\label{Eq_dfxxdt}
 \ee
where $\f{d\sigma}{d^2q}(q^2,\tau)=\alpha_s
C_F\int\f{d^2q}{(2\pi)^2}|V(q^2,\tau)|^2$. The solution of Equ. (\ref{Eq_dfxxdt}) with the initial condition (\ref{Equ:fxx0}) takes the following form
\bea  
f(x_\perp,x_\perp^\prime,\tau,\tau_0)
&=&\varphi(x_\perp)\varphi^*(x_\perp^\prime)\exp\left[\int_{\tau_0}^\tau d\tau \rho(\tau)
\int{d^2q}\f{d\sigma}{d^2q}(q^2,\tau)\left(e^{i\bd{q}\cdot\bd{x}_\perp} + e^{-i\bd{q}\cdot\bd{x}_\perp^\prime}-2\right)\right]\nn\\
&\simeq&\varphi(x_\perp)\varphi^*(x_\perp^\prime)e^{-\f14 \bra
p^2_{\perp}(\tau,\tau_0)\ket(x_\perp^2+x_\perp^{\prime2})},\label{equ:fxxt}
\eea
 where $ \bra p^2_{\perp}(\tau,\tau_0)\ket$ is given by Equ. (\ref{equ:p2t}) and we have made the approximation that $x_\perp$ and $x_\perp^\prime$ are much less than the inverse of the typical momentum transfer.

Now, we are ready to calculate the survival probability of the heavy quarkonium up to time $\tau$, which is given  by 
\be 
P_{sur}(\tau,\tau_0) =\int d^2x_{\perp} d^2x_{\perp}^\prime f(x_\perp,x_\perp^\prime,\tau,\tau_0)f^*(x_\perp,x_\perp^\prime,\tau_0)
.\ee
According to (\ref{equ:fxxt}), we have
\be 
P_{sur}(\tau,\tau_0)=\int d^2x_{\perp} d^2x_{\perp}^\prime
|\varphi(x_\perp)|^2|\varphi(x^\prime_\perp)|^2e^{-\f14 \bra
p^2_{\perp}(\tau,\tau_0)\ket(x_\perp^2+x_\perp^{\prime2})}.\label{equ:Psur}
\ee
Let us approximate the heavy-quarkonium overlap function
with Gaussian wave packet
\be 
|\varphi(x_\perp)|^2 = \f1{\pi
\bra a_B\ket^2}e^{-\f{x_\perp^2}{\bra a_B\ket^2}},\label{equ:gaussian}
\ee 
where $\bra a_B\ket^2 = \f{2}{3}a_B^2$ with $a_B$
the size of the quarkonium, is listed in Table.
\ref{tab:quar}\cite{Satz:2005hx}. Inserting Equ.
(\ref{equ:gaussian}) into Equ. (\ref{equ:Psur}), after some algebra
we obtain

\be 
P_{sur}(\tau,\tau_0) = \f{16}{\left[4 + \bra a_B\ket^2\bra p^2_{\perp}(\tau,\tau_0)\ket\right]^2}.\label{equ:Psur_gau}
\ee

\begin{table}
\caption{Static properties of heavy
quarkonia: in the rest frame of the quarkonium, we take the creation time $\tau_B=1/E_B$ with $E_B$ the binding energy and the typical momentum $p=1/\bra a_B\ket$ with $\bra a_B\ket$ the size of the quarkonium.}\label{tab:quar}\vspace{3mm}
 \centerline{
\begin{tabular}{|c||c|c|c|c|c|}
\hline quarkonia
&mass~(GeV)&$E_B$~(GeV)&$\tau_B$~(fm)&$p$~(GeV)&$a_B$~(fm)\\\hline
\hline $J/\Psi$&3.10&0.64&0.31&0.40&0.5\\\hline
$\Upsilon$&9.46&1.10&0.18&0.71&0.28\\\hline
\end{tabular}}
\end{table}

Equ. (\ref{equ:Psur}) justifies the following physical picture for the dissociation of a heavy quarkonium due to multiple scatterings\cite{Dominguez:2008be}: if the quark and anti-quark in the
quarkonium pick up transverse momenta greater than the typical
momentum in the bound state, that is, $\sqrt{\bra p^2_{\perp}\ket}\gtrsim1/\bra a_B\ket$, the quarkonium breaks
up.  Since the
dominant contribution to $P_{sur}$ comes from the integration of
$x_\perp$ and $x_{\perp}^\prime$ over the region of $x_\perp^2 \lesssim
\bra a_B\ket^2$ and $x_\perp^{\prime2} \lesssim \bra a_B\ket^2$, one can easily see
that
\begin{itemize}
\item if $\bra p^2_{\perp}\ket \gg 1/\bra a_B\ket^2$, $P_{sur}\ll 1$;
\item if $\bra p^2_{\perp}\ket\ll 1/\bra a_B\ket^2$, $P_{sur}\simeq 1$.
\end{itemize}
Such a quantitative behavior analysis tells us that the transition between large and small survival probability must happen at $\bra p^2_{\perp}\ket\sim 1/\bra a_B\ket^2$, and this is exactly the manifest of the above physical picture.

\subsubsection{The multiple scattering of a heavy-quarkonium with $t\gtrsim t_{form}$}

\begin{figure}
\begin{center}
\includegraphics[width=12cm]{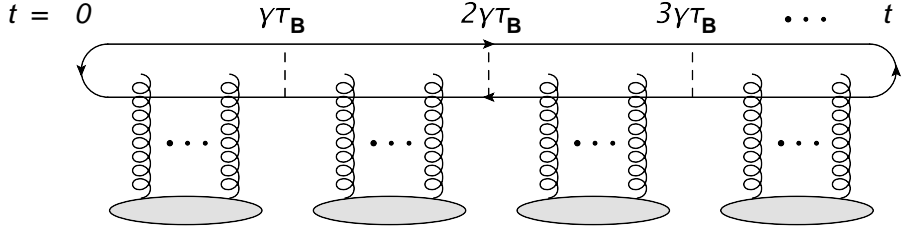}
\end{center}
\caption{The picture of the dissociation of a heavy-quarkonium under multiple scattering within a time interval $t=L\gtrsim t_{form}\equiv \gamma\tau_B=\gamma/E_B$:  the quark(anit-quark) is pulled once by the color field of anti-quark(quark) within a period
$\Delta t\lesssim\gamma\tau_B$. Since the quarkonium can already dissociate with some probability within $\Delta t \sim t_{form}$, as calculated in Equ. (\ref{equ:Psur}), the mutual attraction between the quark and anti-quark is not enough for the binding. And the final-state survival probability is the product of that in each time interval of $\Delta t \sim t_{form}$ within the time interval $t$ under the assumption that the only effect of interaction between the quark and anti-quark within $\Delta t \sim \gamma \tau_B$ is to completely repair the quarkonium..}\label{Fig:ms_beyond_tau}
\end{figure}

In this subsection, we calculate the survival probability of a quarkonium traveling in a slab of expanding QGP within a time $t=L \gtrsim t_{form}$. In this case, one can not neglect the interaction(attraction) between the quark and anit-quark. The wave function of the quarkonium, in the interaction picture, completely changes and the $x_\perp$ coordinate is not "frozen" any more. The complete calculation is beyond the eikonal approximation and we use an approximate picture illustrated in Fig.\ref{Fig:ms_beyond_tau} to estimate the multiple scattering effect.  

In this picture, the quark(anit-quark) is pulled once by the color field of anti-quark(quark) within a period
$\Delta t\sim\gamma\tau_B=\gamma/E_B$.  In the vacuum, the quark(anti-quark) can always be pulled back into the bound state, that is, the quarkonium can completely repair itself. In the medium, because of screened inter-quark potential and the multiple screening effect, the mutual attraction between the quark and anti-quark might not be enough for the binding. Still, we assume that the only effect of interaction between the quark and anti-quark within $\Delta t \sim \gamma \tau_B$ is to completely repair the quarkonium. Obviously, under such an assumption we totally neglect the color screening effect and we will give a more detailed discussion in the Conclusion. The picture could give the parametrically right answer by taking $\Delta t\sim\gamma\tau_B=\gamma/E_B$ even though there is always a constant we can not really control\cite{Dominguez:2008be}. In the following, we take $\Delta t = \xi \gamma \tau_B$ with $\xi$ a constant.  Therefore, the final-state survival probability is the product of that in each time interval of $\xi\gamma\tau_B$ within the time interval $t$, that is,
\be
P_{sur}(\tau, \tau_0) \simeq \left(\prod\limits_{i=1}^n P_{sur}(\tau_i, \tau_{i-1})\right)P_{sur}(\tau, \tau_n),\label{equ:psur_beyond}
\ee
where the time for the quarkonium travels in the medium is $t \leq (n+1)\xi\gamma\tau_B $ and $\tau_i$ is the invariant time corresponding to $i\xi\gamma\tau_B$.

In perturbative
thermal QCD, one has 
\be 
c_s^2 = \f1{3} +\mathcal{O}(\alpha_s^2).
\ee
Keeping only leading terms in
$\mathcal{O}(\alpha_s)$, we have
\be 
\bra p^2_{\perp}(\tau,\tau_0)\ket \simeq \hat{q}_0\tau_0\ln\f{\tau}{\tau_0},\label{equ:qt}
\ee
 Taking the QGP as a weekly coupled gluon gas, $\hat{q}$ and the energy density $\epsilon$ respectively take the following forms\cite{Baier:2006fr}
\be 
\hat{q} =\f{8\zeta(3)}{\pi}\alpha_s^2N_c^2T^3,
\ee 
and
\be 
\epsilon = \f{8\pi^2}{15} T^4,
\ee 
with $\zeta(3)\simeq 1.202.$

In the following sections, we will only take $ \tau_0$ and $\hat{q}_0(T_0)$ as parameters and choose $c_s^2 =\f1{3}$. The transportation coefficient $\hat{q}_0$ characterizes the thermodynamic properties of the background medium at the thermalization time $ \tau_0$. With $ \tau_0$ and $\hat{q}_0$ fixed by RHIC's data about $J/\Psi$ suppression, it is possible for us to decipher some information about thermalization.

\section{$J/\Psi$ suppression at RHIC}\label{sec:RHIC}

\begin{figure}
\begin{center}
\includegraphics[width=10cm]{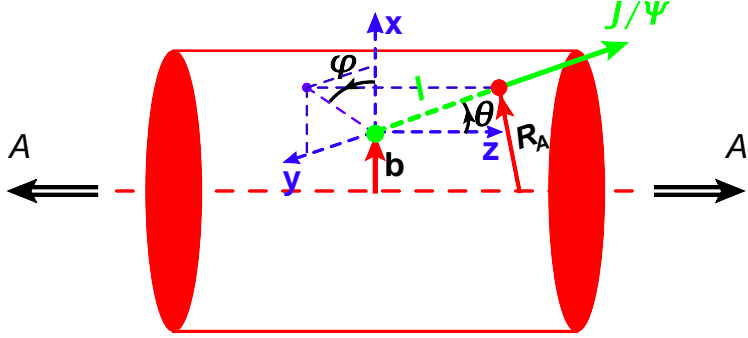}
\end{center}
\caption{Heavy-quarkonia suppression by final-state multiple scatterings in most central Au~+~Au collisions: the background medium is assumed to be thermalized within a short time $\tau_0\sim 1~\mbox{fm}$\cite{Kolb:2001qz,Teaney:2000cw}, which is still larger than $ \tau_B$, the formation time of heavy
quarkonia. After the thermalization time $ \tau_0$, a
quarkonium created at the impact parameter $\bd{b}$ travels in the QGP cylinder with a
radius $R_A$. It may dissociate into uncorrelated $Q\bar{Q}$ pairs due to multiple scatterings  if  $l$, the length of QGP through which it travels, is long enough.}\label{Fig:jpsi}
\end{figure}

In this section we investigate the significance of
final-state multiple scatterings to heavy-quarkonia suppression in a simplified model illustrated in Fig. \ref{Fig:jpsi}. We assume that the background medium is thermalized in a short time $\tau_0\sim 1~\mbox{fm}\ll R_A$, the radius of the QGP cylinder. From Table \ref{tab:quar}\cite{Satz:2005hx}, one can see that the formation time of heavy
quarkonia $ \tau_B$ is shorter than $ \tau_0$. To simplify our discussion, we use the Glauber model\cite{Miller:2007ri} at the stage of heavy-quarkonia creation, that is, the heavy quarkonia are created in nucleon-nucleon binary collisions of the two colliding nuclei at time $\tau_B$. After $ \tau_0$, the
quarkonia travel in the QGP cylinder with a
radius $R_A$. They may dissociate into uncorrelated $Q\bar{Q}$ pairs due to multiple scatterings. In this way, the production of heavy quarkionia is suppressed. In the following, we will calculate the rapidity dependence of the nuclear modification factor $R_{AA}$ in this simplified model.

\subsection{The nuclear geometry}

In most central collisions, a QGP cylinder with a radius $R_A$ is formed after thermalization. We will take $R_A$ to be approximately equal to the radius of the colliding nuclei. Let us introduce the nuclear thickness function $T_A(b)$
 \be 
 T_A(b) = \int dz \rho_A(z,\vec{b}),\label{equ:Ta}
 \ee
where the Woods-Saxon nuclear density $\rho_A$  is given by
\be 
\rho_A(r) =
\f{\rho_{nm}}{1+\exp\left[\left(r-R_A\right)/a\right]},
\ee
with  $\rho_{nm}=0.17~\mbox{fm}^{-3}$, $R_{A} =
6.38~\mbox{fm}$ and $a = 0.53~\mbox{fm}$ for Au\cite{Hahn:1956zz}.
The mean number of binary collisions
in most central collision is defined as
\be
N_{coll}=\int
d^2b\sigma_{pp} T_{A}^2(b),
\ee
with the inelastic total cross-section for $p+p$ collisions $\sigma_{pp}\simeq 40~\mbox{mb}$\cite{Amsler:2008zzb}.

As illustrated in Fig. \ref{Fig:jpsi}, after created in the binary nucleon-nucleon collisions at the impact parameter $\bd{b}$, a heavy quarkonium moves at an angle $\theta$ relative to the cylinder axis. Let us calculate $l$, the length of the background medium through which it travels, i.e., the dashed line protion of the quarkonium's trajectory in Fig. \ref{Fig:jpsi}. It also depends on the azimuthal angle $\varphi$.  Given $(b, \theta, \varphi)$, one can easily get

\be
l(b, \theta, \varphi)=\f{\sqrt{R_A^2 - b^2\sin^2\varphi} - b\cos\varphi}{\sin\theta},\label{Equ:l}
\ee
with $l(b, \theta, \varphi) = 0$ at $b>R_A$.

\subsection{The nuclear geometry}

The nuclear
modification factor $R_{AA}$ of the heavy quarkonium $(\bar{Q}Q)$ is defined as\cite{Adare:2006ns}
\be
 R_{AA} =\frac{dN_{\bar{Q}Q}^{AA}/dy}{N_{coll}dN_{\bar{Q}Q}^{pp}/dy},\label{Equ:r_aa_def}
\ee
with $dN_{\bar{Q}Q}^{AA}/dy$ being the quarkonium $(\bar{Q}Q)$ yield
in A~+~A collisions, $N_{coll}$ the mean number of binary collisions
in the centrality bin, and $dN_{\bar{Q}Q}^{pp}/dy$ the quarkonium
$(\bar{Q}Q)$ yield in $p+p$ collisions. If heavy ion collisions could be taken as uncorrelated nucleon-nucleon binary collisions of the two colliding nuclei, there would be no heavy-quarkonia suppression, i.e., $R_{AA}=1$.  Both the initial cold nuclear
effects, such as gluon saturation effect\cite{Kharzeev:2008cv} and final-state effects\cite{Thews:2005vj,Capella:2007jv}  can make $R_{AA}$ different from 1. In this paper,   we will focus on final-state multiple scattering as discussed in the previous section. With the following observations
\begin{itemize}
\item (a) the quarkonium
$(\bar{Q}Q)$ yield in $p+p$ collisions  $dN_{\bar{Q}Q}^{pp}/d\varphi dy$ is independent of $\varphi$,\vspace{2mm}
\item (b) $T_A(b)$ has rotational symmetry in the impact parameter space,
\end{itemize}
we have
\bea 
R_{AA} &\simeq& P_0 \frac{\int
d^2b\sigma_{pp} T_{A}^2(b)
\int_0^{2\pi}d \varphi d^2N_{\bar{Q}Q}^{pp}/d \varphi dy
P_{sur}(\tau((b, \varphi,y)),\tau_0)}{N_{coll}dN_{\bar{Q}Q}^{pp}/dy}\nn\\
&=& \f{P_0}{N_{coll}}\int dbb\sigma_{pp} T_{A}^2(b)
\int_0^{2\pi}d \varphi P_{sur}(\tau(l(b, \varphi,y),\tau_0)),\label{equ:r_aa} 
\eea 
where  $l$, the length of the QGP through which the heavy quarkonium
travels, is given in Equ. (\ref{Equ:l}) and $P_0$ is the contribution from cold nuclear effects. As a first estimate we will take $P_0$ as the average value of $R_{AA}$ due to cold nuclear effects over rapidity $y$.

One observation about $J/\Psi$ production in $p~+~p$ collisions at $\sqrt{s_{NN}} = 200$~GeV by the PHENIX experiment at RHIC is that the average transverse momentum squared $\bra p^2_T\ket$ at forward rapidity is not significant lower than that at midrapidity\cite{Adare:2006kf}. As a first estimate, we take the typical transverse momentum of $J/\Psi$ produced in binary $p~+~p$ collisions to be independent of rapidity $y$. In this case, one can easily get their typical momentum $p$ and typical velocity $v_0$
\be 
p=\f{p_T}{\sin\theta},\ee and\be v_0(y) = \left(\f{\bra p^2\ket}{m_{J/\Psi}^2 + \bra p^2\ket}\right)^{\f1{2}} =\left(\f{\bra p^2_T\ket}{m_{J/\Psi}^2\sin^2\theta + \bra p^2_T\ket}\right)^{\f1{2}},\ee 
with 
\be 
\sin\theta = \f{2e^y}{1 +
e^{2y}},\label{theta}\ee and \be\bra p^2_T\ket \simeq 3.75~\mbox{GeV}^2.
\ee
Accordingly, the effective length $\tau(l(b,\theta,\phi))$ is given by
\bea
 \tau(l(b,\phi,y))& =& \sqrt{ \left( \f{l}{v_0} \right)^2 - l^2\cos^2\theta }\nn\\
 &=& \left(  \sqrt{R^2_A - b^2\sin^2\varphi} - b\cos\varphi \right) \left( 1 + \f{m_{J/\Psi}^2 }{\bra p^2_T\ket} \right)^{\f1{2}},\label{equ:tau_l}
 \eea
 and
 \bea  
 \tau(\gamma\tau_B)&=&\sqrt{t^2 - z^2(t)} = \sqrt{t^2 - v_0^2 t^2\cos^2\theta}\nn\\
 &=& \sqrt{1 - v_0^2\cos^2\theta}(t_0 + \gamma\tau_B)=\tau_0 + \left(1 + \f{\bra p^2_T\ket}{m_{J/\Psi}^2 }\right)^{\f1{2}} \tau_B,\label{equ:tau_i}
 \eea
where $z(t)$ is the trajectory of the quarkonium. To avoid extending our formula to the non-perturbative regime, we need to define a critical invariant time $\tau_c$, such that,
\be
T(\tau_c) = T_0 \left( \f{\tau_0}{\tau_c} \right)^{c_s^2} = T_c,
\ee
with $T_c$ the critical temperature for the deconfinement phase transition. And we use a modified formula for $R_{AA}$ as follows
\be 
R_{AA} \simeq \f{P_0}{N_{coll}}\int dbb\sigma_{pp} T_{A}^2(b)
\int_0^{2\pi}d \varphi P_{sur} \left( \mbox{min} \left\{ \tau(l(b, \varphi,y)), \tau_c \right\},\tau_0 \right).\label{equ:r_aa_n} 
\ee

\begin{figure}[t]
\begin{center}
\includegraphics[angle=270,width=12cm]{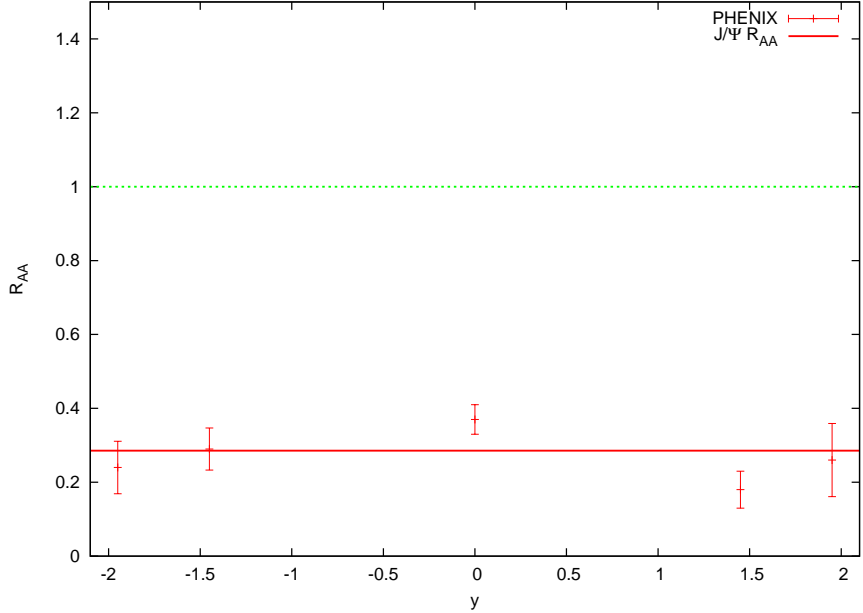}
\end{center}
\caption{Fits about heavy-quarkonia suppression at RHIC: because of boost invariance of the background medium, $R_{AA}$ is independent of $y$. Our task is reduced to fitting the data about $J/\Psi~R_{AA}$ vs $y$ in most central Au + Au collisions in Ref. \cite{Adare:2006ns} to a straight line: $R_{AA}=0.286$ with $\chi^2=9.4$. }\label{fig:r_aa}
\end{figure}

In the following, we are interested in the nuclear modification factor for $J/\Psi$ production $R_{AA}$ only  in the rapidity range $-2.0\lesssim y \lesssim 2.0$. In this case ,we get the Lorentz factor for a $J/\Psi$ with the typical momentum $\gamma\lesssim 1.88$. And this justifies that the formation time $\gamma \tau_B< \tau_0\sim 1~$fm.

\subsection{Discussion of the fits}

\begin{figure}
\input{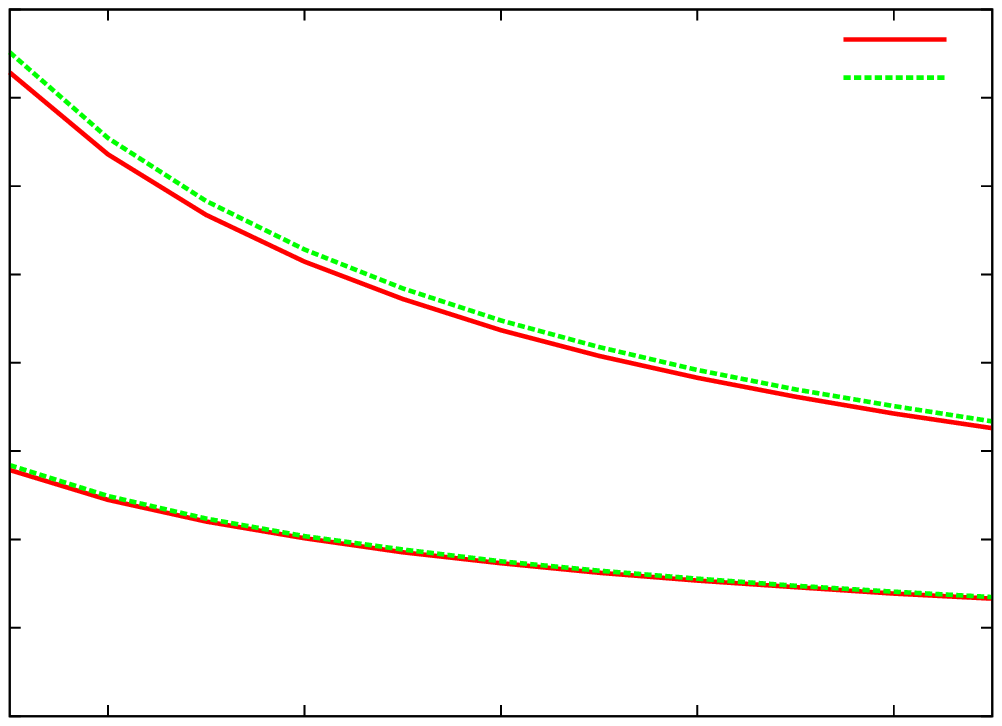}
\caption{The degeneracy of the paramters $\hat{q}_0$ and $\tau_0$: because $R_{AA}$ is independent of $y$, we can not break the degeneracy of $\hat{q}_0$ and $\tau_0$ by only fitting the data about $J/\Psi~R_{AA}$ vs $y$ in most central Au + Au collisions in Ref. \cite{Adare:2006ns}. Given $\tau_0$, $P_0$ and $\xi$, we can decipher the thermodynamical properties at $\tau_0$ by fitting RHIC's data as shown in Fig. \ref{fig:r_aa}. One can see that the transportation coefficient $\hat{q}_0$ dose not sensitively depends on our choice of $\tau_0$ in the range $0.5~\mbox{fm}<\tau_0<1.5~\mbox{fm}$ and $\xi$.}\label{fig:qhat_tau}
\end{figure}

\begin{table}
\caption{Fit results.}\label{tab:fit}\vspace{3mm}
 \centerline{
\begin{tabular}{c c c c c c c c}
\hline\hline
~~~$P_0$~~~&~~~$\xi$~~~&~~~$\alpha_s$~~~&~~~$T_c$~[GeV]~~~&~~~$\tau_0$~[fm]~~~&~~~$T_0$~[GeV] ~~~&~~~$\hat{q}_0~[\mbox{GeV}^2/\mbox{fm}]$~~~&~~~$\epsilon_0$~[GeV/fm$^3$]~~~ \\\hline
\hline
 1.0 	&0.5&0.5 & 0.19 & 0.5~-~1.5 & 0.299~-~0.247  & 0.93~-~0.53 & 5.44~-~2.55 \\
 	&1.0&0.5 & 0.19 & 0.5~-~1.5 & 0.301~-~0.248  & 0.95~-~0.53 & 5.62~-~2.60 \\\hline
 0.4	&	0.5&0.5 & 0.19 & 0.5~-~1.5 & 0.240~-~0.212  & 0.48~-~0.33 & 2.25~-~1.39 \\
 	&1.0&0.5 & 0.19 & 0.5~-~1.5 & 0.240~-~0.212  & 0.48~-~0.33 & 2.28~-~1.40 \\\hline
\end{tabular}}
\end{table}

We use Equ. (\ref{equ:r_aa_n}) and (\ref{equ:psur_beyond}) for our fits and perform numerically the $z$ integration in Equ. (\ref{equ:Ta}) and the $ \varphi$ and $b$ integration in Equ. (\ref{equ:r_aa_n}). According to (\ref{equ:tau_l}), (\ref{equ:tau_i}) and (\ref{equ:r_aa_n}), $R_{AA}$ is independent of rapidity $y$, and this is the manifest of Bjorken boost invariance of the background medium. Finally, our task is reduced to fitting the data about $J/\Psi~R_{AA}$ vs $y$ in most central Au + Au collisions in Ref. \cite{Adare:2006ns} to a straight line. As showed in Fig. \ref{fig:r_aa}, in most central collisions the best fit for $R_{AA}$ versus $y$ is $R_{AA}=0.286$ with $\chi^2=9.4$. 

Our fit results are presented  in Table \ref{tab:fit} and Fig. \ref{fig:qhat_tau}. In Fig. \ref{fig:qhat_tau}, we take $\hat{q}_0$ as a function of $\tau_0$. To see the effects of the cold nuclear matter to our estimates, we show two sets of results with $P_0=1.0$ and $P_0=0.40$ respectively. In the case with $P_0 = 1.0$ we neglect the contribution due to cold nuclear effects. Therefore, we can get the largest estimated value of $\hat{q}_0$ in our estimate. In the other case with $P_0 = 0.4$ we estimate the cold nuclear effects by taking $\bra p_T^2\ket = Q_s^2$ with $Q_s^2 = C_F/N_c 1.25~\mbox{GeV}^2$, the quark saturation momentum squared at RHIC's energy\cite{Mueller:2002kw}.  This gives the lower bound to the value of $\hat{q}_0$ in our analysis. The exact value of $P_0$ at RHIC's energy requires more careful analysis and beyonds the scope of our paper. With $\xi$ and $P_0$ fixed, for each value of $\tau_0$\cite{Kolb:2001qz,Teaney:2000cw}
, we can get a unique value for $\hat{q}_0$ by fitting RHIC's data. One can see that the transportation coefficient $\hat{q}_0$ dose not sensitively depends on our choice of $\tau_0$ in the range $0.5~\mbox{fm}<\tau_0<1.5~\mbox{fm}$ and $\xi$. In Table \ref{tab:fit}, we list the value of each parameter in our formula.  We choose $T_c \simeq 190$~MeV, which is indicated by the lattice simulation\cite{Petreczky:2009at}. Our results of $\hat{q}_0 \simeq 0.33 - 0.95~\mbox{GeV}^2$/fm are in sharp contrast with those obtained by the analysis of high-$p_T$ hadron spectra in Ref. \cite{Eskola:2004cr} in which the authors conclude that the values of the time-averaged  $\hat{q}$ exceed $5~\mbox{GeV}^2$/fm. In Ref. \cite{Arleo:2006xb}, $\hat{q}_0$ is estimated to be  $1.6-2.0~\mbox{GeV}^2/\mbox{fm}$ in central Au+Au collisions at RHIC, which is still at least 2 times larger than our results. Therefore, the puzzle that the $J/\Psi$ suppression is observed to be smaller than expected from the analysis of the energy density at RHIC as discussed in Ref. \cite{Adare:2006ns} also shows up in our calculation.

At the end of this subsection, let us discuss briefly the validity of eikonal approximation within $\Delta t\sim t_{form}$ in describing $J/\Psi$ production at RHIC energy. Our calculation in Sec. \ref{Sur} is justified only if $q \ll 2p_c \simeq p_{J/\Psi}$ where  $p$ is the typical momentum of the (anti-)heavy quark and $q$ is the typical momentum transfered under each individual scattering between the (anti-)quark and one constituent of the QGP.  In a thermal medium, one has $q\sim T\lesssim T_0 \simeq 0.25~\mbox{GeV}$. At RHIC energy, $2p_c \simeq p_{J/\Psi}\geq\sqrt{\bra p^2_T\ket}\simeq 1.936~\mbox{GeV}$. As a result, $\f{q}{2p_c}\lesssim 0.13$. Therefore, at RHIC energy, the eikonal approximation should be good enough for us to estimate the significance of the final-state multiple scattering to heavy-quarkonia suppression within $\Delta t\sim t_{form}$.

\section{Conclusion}\label{sec:con}

Due to their small size, heavy quarkonia act as an excellent hard probe to the properties of the background QCD matter in heavy ion collisions. In this paper, we focus on their another static property, the short formation time $\tau_B$. If the background medium is thermalized at time $\tau_0\gtrsim \tau_B$, one can use heavy quarkonia to probe the thermodynamic properties at time $\tau_0$. To illustrate our point, we present a simplified model in Fig. \ref{Fig:jpsi} to discuss the dissociation of heavy quarkonia due to final-state multiple scatterings in Bjorken's expanding QGP. From fit results of $J/\Psi~R_{AA}$ versus $y$ in most central Au~+~Au collisions at $\sqrt{s_{NN}}=200~\mbox{GeV}$ at RHIC, we get the transportation coefficient $\hat{q}_0 \simeq (0.33-0.95)~\mbox{GeV}^2$/fm (accordingly the energy density $\epsilon_0\simeq(1.39-5.62)~\mbox{GeV}/\mbox{fm}^3$ in perturbative thermal QCD). Our results about the transportation coefficient $\hat{q}_0$ are in sharp contrast with those obtained by the analysis of high-$p_T$ hadron spectra in Ref. \cite{Eskola:2004cr}.

\begin{figure}
\begin{center}
\includegraphics[width=12cm]{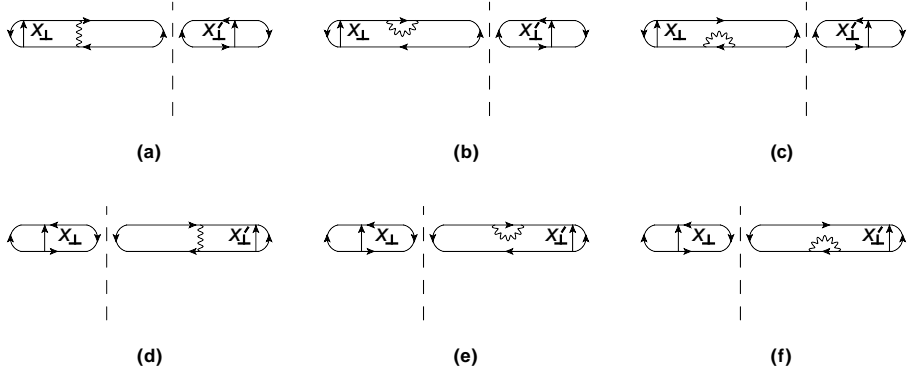}
\end{center}
\caption{The attraction between the quark and anti-quark: $\bd{x}_\perp$ and
$\bd{x}_\perp^\prime$ are respectively the relative transverse coordinates of the quark and antiquark in the amplitude and in the conjugate amplitude. The quark and anti-quark interact with each other via thermal gluons with a mass $\mu_D$.}\label{Fig:one_attraction}
\end{figure}

In this work, we use Bjorken's boost-invariant hydrodynamical model to describe the evolution of the background medium and obtain a rapidity-independent $R_{AA}$. Obviously this is one reason for the deviation of the measured $J/\Psi~R_{AA}$ at RHIC from such a rapidity-independent behavior just as the deviation from the central-plateau structure of particle multiplicity in the rapidity spectra observed at RHIC. One refinement of this work is to introduce the centrality dependence of $R_{AA}$ by using more complicated hydrodynamic model to describe the background medium. Another refinement of this work is to investigate in details the competition between the multiple scattering effect and the attraction between the quark and antiquark in a quarkonium within a time interval of $\Delta t\sim R_A\gg \tau_B$. The inclusion of the diagrams shown in  Fig. \ref{Fig:one_attraction} in Equ. (\ref{Eq_dfxxdt}) enables one to compare the multiple scattering effect, Debye screening and the melting if the imaginary part of $\mu_D^2$ included\cite{Laine:2008cf}.
Besides, other possibly important effects have not been addressed, including the gluon saturation effects\cite{Kharzeev:2008cv} and other final-state effects, e.g.,  jet quenching\cite{Wiedemann:2009sh}, the nuclear absorption and recombination effects\cite{Thews:2005vj,Capella:2007jv}. Moreover, in this paper  we only discuss the lowest $c\bar{c}$ bound states. In a more quantitative analysis, one should also include the contribution from excited quarkonium states to ground-state production. Accordingly, still another refinement of this work is to include all other effects mentioned above to make a quantitative comparison. 

\section{Acknowledgements}
We are greatly indebted to Prof. A. H. Mueller for numerous
illuminating discussions and comments. We would also like to thank Prof. R. Baier, Prof. M. Laine,
Dr. C. Marquet, Dr. A. Vuorinen and Prof. Y. Mao for useful discussions. This work is
supported by the National Natural Science Foundation of China (Nos.
10721063, 10975003). B.W. also acknowledges partial financial support from the Humboldt foundation
through its Sofja Kovalevskaja program.

\end{document}